\documentclass[aps,prb,twocolumn,groupedaddress,showpacs]{revtex4-1}
\usepackage{mathbbol,amsmath,amsfonts,amssymb,bm,color,dsfont}
\usepackage{graphicx,psfrag,array}

\usepackage{amsmath}
\usepackage{mathtools}
\usepackage{braket}
\usepackage{graphicx,psfrag,color}
\usepackage{dcolumn}
\usepackage{bm}
\usepackage{mathbbol, appendix}
\usepackage{multirow}
\usepackage{color,soul}

\providecommand{\ignore}[1]{}
\providecommand{\aucmnt}[1]{#1}
\def\be{\begin{equation}}
\def\ee{\end{equation}}
\renewcommand{\aucmnt}[1]{}

\newcommand{\Comment}[1]{}

\newcommand{\Eq}[1]{Eq.~(\ref{#1})}

\begin{document}

\title{Thin torus perturbative analysis of elementary excitations in the Gaffnian and Haldane-Rezayi quantum Hall states}
\author{Amila Weerasinghe}
\author{Alexander Seidel}
\affiliation{Department of Physics, Washington University, St.
Louis, Missouri 63130, USA}

\date{\today}
\begin{abstract}
We present a systematic perturbative approach to study excitations in the thin cylinder/torus limit of the quantum Hall states. The approach is applied to the Haldane-Rezayi and Gaffnian quantum Hall states, which are both expected to have gapless excitations in the usual two-dimensional thermodynamic limit. For the Haldane-Rezayi state, we confirm that gapless excitations are present also in the ``one-dimensional'' thermodynamic limit of an infinite thin cylinder, in agreement with earlier considerations based on the wave functions alone.
In contrast, we identify the lowest excitations of the Gaffnian state in the thin cylinder limit, and conclude that they are gapped, using a combination of perturbative and numerical means.
We discuss possible scenarios for the cross-over between the two-dimensional and the one-dimensional thermodynamic limit in this case.
\end{abstract}
\maketitle

\section{Introduction}

Quantum Hall states represent a prime example of phases of matter
for which ideas of dimensional reduction are of central importance.
This is rooted in the bulk-edge correspondence for topological phases 
described by Chern-Simons quantum field theories.\cite{witten}
This correspondence is also manifest in certain preferred or ``special''
microscopic trial wave functions used to describe quantum Hall phases,
and whose analytic structure is that of conformal blocks of the unitary rational
conformal field theory (CFT) describing the edge of the same phase.\cite{MR}
This situation extends to trial wave functions whose analytic structure is derived from
conformal blocks in nonunitary CFTs. Examples of the latter kind are the Gaffnian
state\cite{gaffnian} and the {Haldane-Rezayi}\cite{HR} state.
Here, the physical interpretation of this correspondence is more subtle,
as the respective nonunitary CFT is not acceptable as the description of a physical edge.
In these cases, it has been argued\cite{gaffnian,NickRead1,NickRead2} that a local microscopic
Hamiltonian stabilizing such a wave function as ground state must have gapless excitations.
In other words, such wave functions are not expected to describe topological (gapped) phases. 
This conjecture has stimulated numerous theoretical and numerical investigations,\cite{NickRead2,gaffnian,ReadGreen,Wang,Entangle,bernevig2012screening, jolicoeur}
though providing direct evidence and/or microscopic characterization of the gapless excitations remains an interesting problem.
In the case of the Haldane-Rezayi state, some insight has been obtained by analyzing
a thin torus (TT) -- or thin cylinder -- limit.\cite{seidel_yang} The TT limit is yet another way to achieve 
a two-dimensional -- one-dimensional (2D--1D) correspondence  in the context of quantum Hall systems.\cite{bergholtz05, Seidel_Moore,Seidel_Lee,Bergholtz,PfaffianTT, Seidel10} In Ref. \onlinecite{seidel_yang}, the very knowledge of the TT limit of the Haldane-Rezayi (HR) wave functions was used to argue that charge-neutral gapless excitations must exist in the TT limit, and the latter have been characterized as certain extended equal-amplitude superpositions of defects (see below). In that argument, the detailed form of the HR parent Hamiltonian was not used, merely the knowledge that it exists and that it has  a zero-energy ground state. 
In this paper, we will show how the features inferred in Ref. \onlinecite{seidel_yang}
can be straightforwardly derived in a perturbative framework, which, as a byproduct,  also reveals the proper dependence of the quadratic dispersion on the (thin) cylinder radius.
As we will review below, it has been cautioned in Ref. \onlinecite{seidel_yang} that while the finding of gapless excitations
in the thin torus limit is quite plausible evidence for their existence in the 2D thermodynamic limit, the converse is not necessarily true.
Indeed, we apply the same perturbative scheme to the Gaffnian state, and find conclusive analytical and numerical evidence that gapless excitations are absent {\em in the TT limit}. We give an asymptotic formula describing the gap where first the thermodynamic limit is taken in one of two spatial directions and then the TT limit is taken in the other direction. As we discuss in detail in Sec. \ref{discussion},
this does not preclude the existence of gapless excitations in the usual 2D thermodynamic limit, though unfortunately, we cannot say more about this from a TT point of view. Nonetheless, we hope that our investigation will shine interesting light on the different possible relations between various types of quantum Hall states and their TT limits.

\section{Gapless Excitations in the Haldane-Rezayi state} \label{HR}

It has been argued in previous studies\cite{seidel_yang} that in the TT limit, the gapless character of the Haldane-Rezayi state is manifest in the limiting forms of the associated wave functions.  Below we develop a perturbative framework that makes these claims explicit. We focus on the top state in the HR sequence with fermionic filling fraction $ \nu=1/2 $.

\begin{figure}[htb]
\includegraphics[width=0.6\columnwidth]{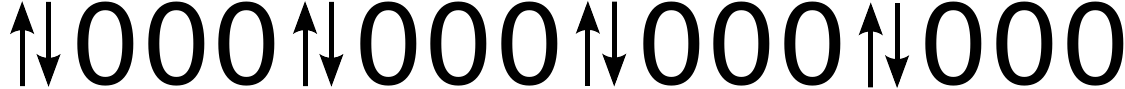}\\
\includegraphics[width=0.6\columnwidth]{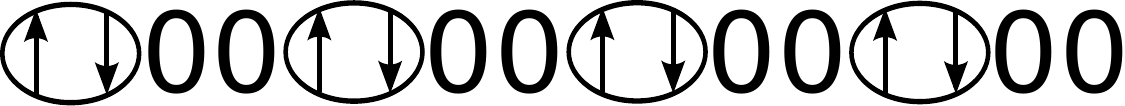}
 \caption{Haldane-Rezayi thin torus ground state patterns, in the usual occupation number
 representation. Zeros denote empty orbitals. The configuration $\uparrow\downarrow$ 
 denotes an up-spin and a down-spin particle occupying the same orbital.
  Ovals denote spin singlets. }
 \label{HRgroundstates.fig}
\end{figure}

\begin{figure}[htb]
\includegraphics[width=0.8\columnwidth]{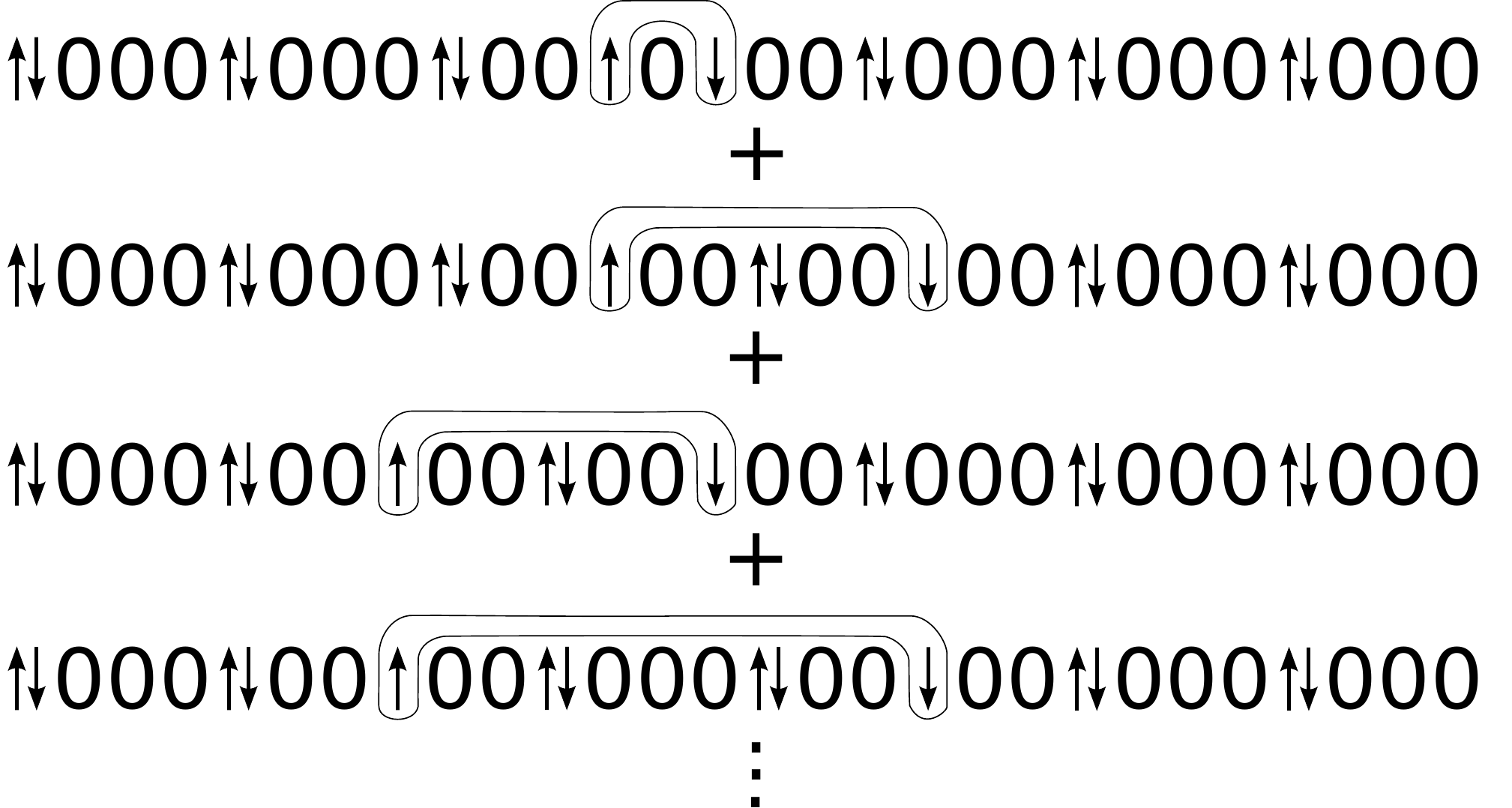}
 \caption{The thin torus limit of a particular ground state of the hollow-core Hamiltonian. The limiting form is an equal-amplitude superposition of states with a delocalized pair of charge-neutral defects forming a singlet. }
 \label{Aprime.fig}
\end{figure}

The two-component HR state is tenfold degenerate\cite{wen_paper} on the torus, with eight of ten ground states 
approaching one of two patterns in the TT limit, up to translations, given in Fig.\ref{HRgroundstates.fig}.
Because of the translational symmetry, the two states shown in the figure account for eight ground states. 
Note that we refer to the two components of fermions as spin-up and spin-down here and in the following.
There are two other special ground states whose TT limits are not fully described by a simple unit cell. These states are in fact  closely related to the presence of  gapless excitations in HR state. One of these special thin torus HR ground state patterns is given in Fig.\ref{Aprime.fig}, which can be understood as a delocalized singlet immersed into and separating two ground states of the first kind in
Fig.\ref{HRgroundstates.fig}. An explicit calculation using perturbation theory will be given explaining how these excitations acquire zero energy.
     
The HR state is known to be the exact-zero energy ground state of the ``hollow-core'' Hamiltonian.\cite{HR}
This is just the $V_1$ Haldane pseudopotential,\cite{haldane_hierarchy}  acting between any two
electrons regardless of their spin. The name ``hollow core'' is alluding to the fact that a $V_0$-term
is allowed between electrons of opposite spin, but is absent in the Hamiltonian.
Here we will work mostly on an infinite cylinder with finite circumference $L_y=2\pi/\kappa$, with
$\kappa$ being the inverse radius. A basis for the lowest Landau level in this geometry is naturally
given by a set of translationally related orbitals $\phi_r$, labeled by an integer $r$, such that the 
$x$ component of the guiding center is well-defined and equal to $\kappa r$, and we set the magnetic length equal to $1$.
In second-quantized form, the Hamiltonian then takes the form of a spin-SU(2) invariant
two-body operator,
\begin{eqnarray}
H=\frac{1}{2}\sum_{m',n',m,n,\alpha,\beta}V_{m'n'mn}\, C_{m',\alpha}^\dag\, C_{n',\beta}^\dag\,  C_{n,\beta}\, C_{m,\alpha}\,,
\end{eqnarray}
where $C_r$ destroys a particle in the state $\phi_r$, and the matrix element
$V_{m'n'mn}$ does not depend on spin indices $\alpha$, $\beta$ because of $SU(2)$ invariance.
This matrix element is therefore just the same as for the $V_1$ pseudopotential in a spin-polarized setting, on a cylinder of radius $1/\kappa$,
which is standard in the literature (see, e.g., Ref. \onlinecite{lee_leinaas}). This gives (with arbitrary but $\kappa$-independent overall normalization)
\begin{eqnarray}\label{HR Hamilt}
H&=&\sum\limits_R\sum\limits_{\alpha,\beta}\sum_{\substack{m'+n'=2R \\ m+n=2R}} \kappa^3 (m-n)(m'-n')\nonumber\\
&\times& e^{-\kappa^2[(m-n)^2+(m'-n')^2]/4}\, C_{m',\alpha}^\dag\, C_{n',\beta}^\dag\,  C_{n,\beta}\, C_{m,\alpha}.
\end{eqnarray}

Now that spin-1/2 degrees of freedom are present in the problem, it should be noted that
the pair interaction defined by \Eq{HR Hamilt} still only acts on triplet pairs. 
This is natural since, in the infinite plane, the $V_1$ pseudopotential is defined 
as a two-particle projection operator projecting on states with relative orbital angular momentum
$1$. No pair forming a spin singlet can have this relative angular momentum.
On the cylinder/torus geometry, however, relative angular momentum is not well defined.
Hence it is worth noting that the fact remains that the interaction annihilates any singlet pair.
This follows already from the fact that the matrix element $V_{m'n'mn}$ is antisymmetric
in $m$ and $n$ (as well as their primed counterparts) whereas any two-particle singlet 
state must have a symmetric orbital part.

We now use the second-quantized Hamiltonian \eqref{HR Hamilt} to set up a perturbative scheme
designed to calculate energies and states in powers of $x=e^{-\frac 12 \kappa^2}$.
To this end we write the Hamiltonian as
\be\label{lambda}
    H=H_0+\lambda H_1\,,
\ee
where $\lambda=1$ is a formal parameter.
$H_0$ contains all terms in the Hamiltonian \eqref{HR Hamilt} 
that are diagonal in the {\em orbital} indices; that is, all terms 
for which the unordered pairs $(m,n)$ and $(m',n')$ are equal,
whereas spin indices may or may not be equal.
$H_1$ contains all the remaining, off-diagonal terms.
We will perform a double expansion. The first of these is the formal
expansion in the parameter $\lambda$. It turns out that each order in $\lambda$
receives multiple contributions (infinitely many, for infinite system size)
in the different powers of the parameter $x$. At any fixed order in $\lambda$,
we will therefore retain only those orders of $x$ that we are interested in.
We claim that in this way, to get all contributions of a certain order $x^\ell$ exactly,
one needs to go only to a certain finite order in $\lambda$, which will depend on
$\ell$. We will not attempt a formal proof of this statement, but it will become
quite apparent that for higher and higher orders in $\lambda$, the leading order
in $x$ will grow systematically.
In our case, we will be interested in terms up to 12th order in $x$, for which
second-order perturbation theory in $\lambda$ will be sufficient.

\begin{figure}[htb]
\includegraphics[width=0.6\columnwidth]{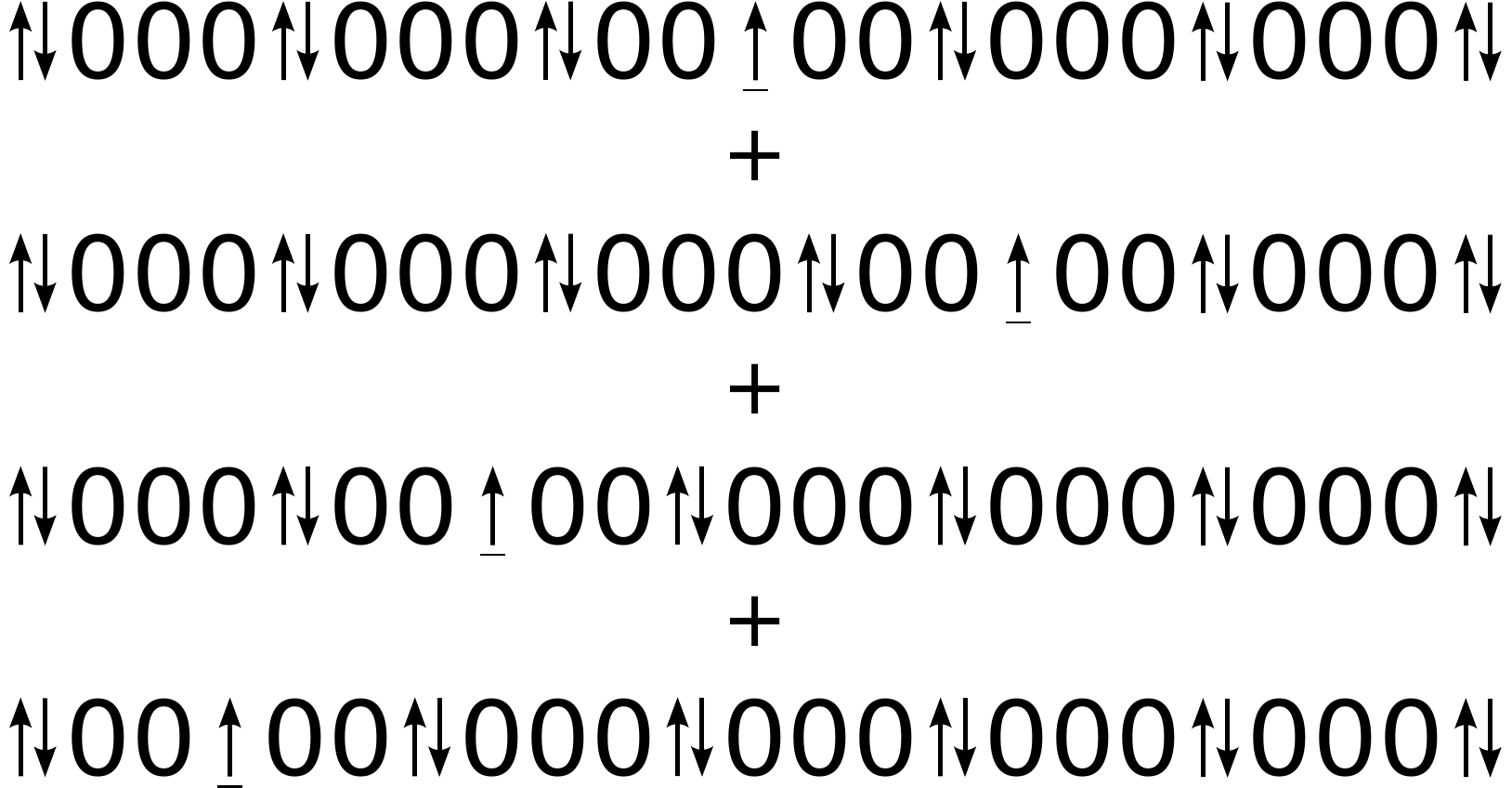}
 \caption{A spin-1/2 defect becomes delocalized in an $ A $-type ground-state pattern}
 \label{onedefect.fig}
\end{figure}

We will first focus on the odd-particle-number sector, for which one has two
degenerate ground-state doublets on the torus.\cite{read_paper}
The relevant thin torus states are discussed in Ref. \onlinecite{seidel_yang}.
They are obtained as a superposition of states of the form shown in Fig.\ref{onedefect.fig},
where a single spin-1/2 defect becomes delocalized in a ground-state pattern of the 
$A$ type (the first of the ground-state patterns in Fig. \ref{HRgroundstates.fig}). 

Clearly, all states contributing to this superposition are degenerate for $H_0$,
and hence we must apply degenerate perturbation theory in $\lambda$.
The leading nontrivial order in $x$ turns out to be $x^{12}$, and we shall be content
with this order. For this, it turns out to go to second order in $\lambda$.
Order-$\lambda^0$ diagonal matrix elements are dominated by the interaction
of the spin-1/2 defect with two neighboring singlets at distance $3$.
It is easy to see from \Eq{HR Hamilt} that each ``bond'' between a spin-1/2
defect and one such neighboring singlet costs an energy of 
\be
  E_0=54x^9\,.
\ee

We shall now consider corrections up to order $x^{12}$ arising in second-order
degenerate perturbation theory in $\lambda$.
For simplicity, we will first consider a three particle system. 
The two ($H_0$)-degenerate thin cylinder ground states are
\begin{eqnarray}
\ket{\Omega_1}&=& C_{0,\uparrow}^\dag C_{3,\downarrow}^\dag C_{3,\uparrow}^\dag\\
\ket{\Omega_2}&=& C_{1,\downarrow}^\dag C_{1,\uparrow}^\dag C_{4,\uparrow}^\dag  
\end{eqnarray}
We also truncate the Hilbert space to consist of five orbitals, $r=0\dotsc 5$, limiting ourselves
to the following two excited states:
\begin{eqnarray}
\ket{\Psi_1}&=& \frac{1}{\sqrt{2}}(C_{1,\uparrow}^\dag C_{2,\uparrow}^\dag C_{3,\downarrow}^\dag - C_{1,\downarrow}^\dag C_{2,\uparrow}^\dag C_{3,\uparrow}^\dag)\\
\ket{\Psi_2}&=& \frac{1}{\sqrt{6}}(C_{1,\uparrow}^\dag C_{2,\uparrow}^\dag C_{3,\downarrow}^\dag -2 C_{1,\uparrow}^\dag C_{2,\downarrow}^\dag C_{3,\uparrow}^\dag+ C_{1,\downarrow}^\dag C_{2,\uparrow}^\dag C_{3,\uparrow}^\dag) 
\end{eqnarray}
Using spin-rotational and other symmetries, these are the only two states in the truncated
Hilbert space that our unperturbed states can mix with. 
Note that they are both $H_0$ eigenstates, though not of the same energy, having energies
$6x$ and $2x+96x^4$ respectively, owing to spin fluctuations that are kept in $H_0$.
In second-order degenerate perturbation theory, one then has the following correction
to the diagonal matrix element:
\begin{eqnarray}
E_{diag}^{(2)}&=& \dfrac{|\braket{\Psi_1|H_1|\Omega_1}|^2}{0-E_{\Psi_1}^{(0)}}+\dfrac{|\braket{\Psi_2|H_1|\Omega_1}|^2}{0-E_{\Psi_2}^{(0)}}\nonumber\\
&=& -54\, x^9 + 1296\, x^{12}- \mathcal{O}(x^{15})\,,
\end{eqnarray} 
where we have only kept terms up to order $x^{12}$.
This reduces these diagonal matrix elements to an energy of order $x^{12}$, which we denote by $V$,
\begin{eqnarray}
V &=& E_0 + E_{diag}^{(2)} \\
V &=& 1296\, x^{12}- \mathcal{O}(x^{15})\,.
\end{eqnarray}
Similarly, the off-diagonal matrix element is obtained as
\begin{eqnarray}
-t&=& E_{off-diag}^{(2)}\nonumber\\
&=&\dfrac{\braket{\Omega_2|H_1|\Psi_1}\braket{\Psi_1|H_1|\Omega_1}}{0-E_{\Psi_1}^{(0)}}+\dfrac{\braket{\Omega_2|H_1|\Psi_2}\braket{\Psi_2|H_1|\Omega_1}}{0-E_{\Psi_2}^{(0)}}\nonumber\\
&=&- 1296\, x^{12}+ \mathcal{O}(x^{15}).
\end{eqnarray}
At order $x^{12}$, we thus have $t=V$, as was correctly inferred from less direct arguments
in Ref. \onlinecite{seidel_yang}.
At order $x^{12}$, we thus obtain the effective Hamiltonian
\be
H_{eff} =
 \begin{pmatrix}
  V & -V  \\
  -V & V
 \end{pmatrix}\,,
 \ee
leading to a single zero-energy eigenstate, which is the equal amplitude superposition of
$\ket{\Omega_1}$ and $\ket{\Omega_2}$, as expected. The energy of the other member
of the formerly degenerate pair is $E_{ext}=2V$.

One may ask if our perturbative scheme is valid, since both zeroth-order and second-order 
matrix elements in $\lambda$ were of the same order $x^9$ in $x$.
At least for the three-particle problem, this question can be settled exactly.
The full Hamiltonian in this truncated Hilbert space corresponds to the 4x4
matrix
\[ \left( \begin{array}{cccc}
54 x^9  & -\frac{18}{\sqrt{2}} x^5 & -\frac{18}{\sqrt{6}} x^5 & 0 \\ [8pt]
-\frac{18}{\sqrt{2}} x^5 & 6 x & 0 & -\frac{18}{\sqrt{2}} x^5 \\[8pt]
-\frac{18}{\sqrt{6}} x^5 &  0 & 2 x+ 96 x^4 & \frac{18}{\sqrt{6}} x^5\\[8pt]
0 & -\frac{18}{\sqrt{2}} x^5 & \frac{18}{\sqrt{6}} x^5 & 54 x^9 \end{array} \right).\]\\
It can be shown exactly that this matrix has one  lowest eigenvalue at zero,
with the next higher up eigenvalue being
\begin{eqnarray}
 E_{ext}&=&\,x + 48 x^4+27 x^9 \\
  &-& x \sqrt{1+3 x^3(32+768 x^3+18 x^5 -864 x^8+ 243 x^{13})}\nonumber
\end{eqnarray}
Expanding the above up to order $x^{12}$, one finds $E_{ext}=2V$ in agreement with our perturbative
approach.
Higher orders in $\lambda$ will thus only contribute subdominant terms in $x$.

\begin{figure}[htb]
\includegraphics[width=0.5\columnwidth]{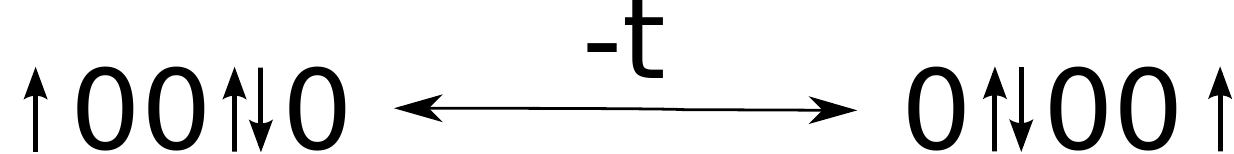}
 \caption{The off-diagonal matrix element delocalizing spin-1/2 defects.}
 \label{offdiag.fig}
\end{figure}

Turning to the $N$-particle problem defined by the Hamiltonian \eqref{lambda} and 
the $H_0$-degenerate subspace described in Fig. \ref{onedefect.fig},
we have, first of all, contributions to the effective Hamiltonian $H_{eff}$ that are exactly analogous
to those in the three-particle problem discussed first. We still find no other processes, at second \underline{or} higher oder in $\lambda$, that contribute to order $x^{12}$ or less in $x$.
Therefore, the picture is similar to the three-particle problem. At order $x^{12}$, each state in 
Fig. \ref{onedefect.fig} has a diagonal energy of $2V$ ($ V $ for each neighboring singlet of the defect).
On top of that, we have a hopping matrix element of the form shown in Fig. \ref{offdiag.fig}, with 
$t=V$.
The defect thus acquires a gapless quadratic dispersion of the form
\be\label{dispersion}
 E(k)=2V-2V\cos(k)\,,
\ee
as predicted in Ref. \onlinecite{seidel_yang}
with $V= 1296\, x^{12}+\mathcal{O}(x^{13})$.
The state corresponding to $k=0$ is the zero-energy ground state corresponding, in the TT limit, to the equal-amplitude superposition of the states shown in Fig. \ref{onedefect.fig}.

Next we discuss the case of even particle number. In this case, the relevant $H_0$-degenerate
subspace is given by all states of the form indicated in Fig. \ref{Aprime.fig}.
Diagonal energies are now of the form $4V$ (except for states such as the first shown in the figure; see below), and we still have the effective defect hopping
shown in Fig. \ref{offdiag.fig}, with $t=V$.
It is found that the equal-amplitude superposition of Fig. \ref{Aprime.fig} still gives a zero-energy state.
The only additional subtlety arises from configurations where the two defects are in closest proximity,
as the first shown in the figure. It was conjectured in Ref. \onlinecite{seidel_yang}
that there should be no energy associated with the two neighboring defects, as long as the latter are forming a singlet. We have already discussed above why this is indeed the case, as the Hamiltonian only acts on triplet pairs. The diagonal energy of such configurations is thus $2V$, and this is exactly required to satisfy the ``detailed balance'' condition giving a zero-energy state.
Moreover, it is clear on variational grounds that boosting the momentum of the delocalized pair would give rise to orthogonal states of arbitrarily small energy, in the thermodynamic limit. Indeed, using the matrix elements discussed here, it is easy to
generalize \Eq{dispersion} to the problem of two defects of ``rapidities,'' $ k_1$ and $k_2$, respectively, for which there will be an eigenstate
of energy $E(k_1)+E(k_2)$, with $E(k)$ as given by \Eq{dispersion}.

We have thus verified all of the conjectures made in Ref. \onlinecite{seidel_yang}, going up to orders
$x^{12}$ in a perturbative framework. At higher order in $x$, we expect that while corrections will be nontrivial, 
the observed ``detailed balance'' between diagonal and off-diagonal terms observed at the present order will continue to hold
and lead to the presence of a zero mode. This must, of course, be true from the fact that 
first-quantized zero mode wave functions can be given exactly that do have delocalized
defects of the form presented here in the TT limit. The arguments for gapless excitations
given in Ref. \onlinecite{seidel_yang}, which we have verified explicitly at the lowest nontrivial order
in perturbation theory here, are expected to hold to all orders. 
For completeness, we briefly repeat and sharpen these arguments in the following.
First of all, in the sector that has only a single defect, one would still find
a tight-binding-type dispersion for this defect, with longer-ranged but exponentially decaying hopping terms. This dispersion must have its bottom edge precisely at zero to
any order in perturbation theory, as we explained above.
Hence, for the single-defect sector (which requires particle number to be odd),
the gapless character of the TT limit follows.

In the even-particle-number case, there must be two defects, which, at higher
orders in perturbation theory, may interact in more complicated ways than
found at the present order. We do know, however, that despite this interaction,
there will be a zero mode to all orders in perturbation theory, featuring two
delocalized defects forming a singlet. Again, this is known since the wave functions
having this behavior can be given exactly. Moreover, the interaction between
these defects will be exponentially decaying. 
For these two degrees of freedom in an infinite system,
general results in scattering theory\cite{reed_simon_vol_III} imply that such local interactions do not change the absolutely continuous 
spectrum of the theory. If we adiabatically switch off the interactions, we are back to the
spectrum associated to two single defects, which is continuous and has its bottom edge at zero,
as discussed above. Hence this is also true in the presence of the interaction. Moreover,
we know that the interaction cannot cause any bound states to appear below zero energy, since the 
Hamiltonian is known to be positive. Thus, the observation that the spectrum
has a continuum above its lowest value at zero
 for the sector containing two delocalized singlet defects
must continue to hold at any order in perturbation theory. The only difference
at the order given here is that the corresponding eigenstates can be worked out easily from the given matrix elements.



\section{Thin torus elementary excitations in the Gaffnian state.}\label{Gaffnian}

\subsection{General considerations\label{gaff}}
The Gaffnian wavefunction is a state of particles at filling factor $ \nu=2/3 (2/5) $ for bosons(fermions).\cite{gaffnian}
Its parent Hamiltonian is a three-particle interaction and has been extensively discussed.\cite{gaffnian, simon2007generalized} 
We will focus on the bosonic case here for simplicity.
On the torus, the ground state is sixfold degenerate, with thin torus states approaching the 
patterns $200200200\dotsc$, $110110110\dotsc$, including translations.
We wish to investigate if a scenario similar to that of the HR state is realized,
and gapless excitations can be identified in the TT limit.

One main difference between the Gaffnian and HR case is the fact that none of the
Gaffnian ground states look ``suspicious'' in the TT limit, whereas the HR state has ground states
(among others) whose TT limit is the equal-amplitude superposition shown in 
Fig. \ref{Aprime.fig}. From the latter, all features derived in the preceding section have 
been correctly inferred previously.\cite{seidel_yang}
Here we investigate a scenario that could nonetheless explain the existence of Gaffnian
gapless excitations
of a similar flavor to those discussed for the HR. Unfortunately, we find that details of this scenario do not work out, and the excitations we discuss are gapped in the TT limit.
However, from comparison with numerics, we do find that these excitations 
are indeed the lowest-energy excitations in the TT limit, and hence the TT limit
of the Gaffnian state is gapped.

A class of parent Hamiltonians for the Gaffnian state can be written as\cite{gaffnian}
\begin{eqnarray}\label{Hamilt}
H&=&\,  V_0 P_{3}^{0}\,+V_2 P_{3}^{2}\,,
\end{eqnarray}
where $V_0$ and $V_2$ are positive constants, and $P_3^0$ and $P_3^2$ are three-particle projection operators that project onto the subspace of relative angular momentum $0$ and $2$, respectively.
Using the results of Ref. \onlinecite{Qi}, this interaction is readily presented in second-quantized form,

\begin{equation}\label{second quantized}
\begin{split}
H=& \sum\limits_{R}\,(q_{R}^\dag\, q_{R}+ C\, Q_R^\dag Q_R)\\
&\text{Where,}\\
 Q_{R}=&\sum_{\substack{\tiny m+n+l=\\3R }}\:\Big\{1-\kappa^2 \Big[(R-m)^2+(R-n)^2+(R-l)^2\Big]\Big\}\\ &\times e^{-\kappa^2\Big[\frac{(R-m)^2+(R-n)^2+(R-l)^2}{2}\Big]} C_n C_m C_l\\
q_R =& \sum_{\substack{\tiny m+n+l=\\3R }}e^{-\kappa^2\Big[\frac{(R-m)^2+(R-n)^2+(R-l)^2}{2}\Big]} C_n C_m C_l\,,
\end{split}
\end{equation}
and where $C>0$ is a constant that controls the relative strength between the two terms in \Eq{Hamilt},
and we have chosen an overall normalization.\footnote{Here, we omit an overall factor $\kappa^2$, that is also not present in Ref. \cite{Qi}.
It needs to be included if a $\kappa$-independent normalization of pseudo-potentials is desired, but is of no consequence in the present context.}
In \Eq{second quantized}, the summation over $R$ is over all values such that $3R$ is an integer.

We now wish to investigate a possible scenario for gapless neutral excitations similar to those of the HR state in the TT limit. Charge neutrality is a key aspect of the domain-wall-type defects studied in the preceding section. Only a neutral defect is necessarily delocalized in the manner seen there, allowing for the gapless character. Charged defects would be subject to greater constraints from ``center-of-mass conservation'' \cite{Seidel_Moore} (momentum conservation around the cylinder axis).
A natural neutral defect between two different Gaffnian TT ground-state patterns is given by the following configuration:
\be\label{defect}
  \dotsc 200200200201011011011011011 \dotsc
\ee
The fact that the above defect is charge neutral can be seen as follows. Starting with the
$200200\dotso$ ground-state pattern, we may consider a pair of particles occupying the same orbital and move one member of the pair to the left neighboring orbital, and the other to the right.
We then obtain \Eq{defect}
by proceeding in this way with double occupancy to the right of the original one.
Such local rearrangement of charge cannot lead to a charged defect.
 As written, the defect should cost a finite energy, as it violates the 
Gaffnian ``generalized Pauli principle'' \cite{bernevig28,bernevig29} of having no more
than two particles in any three adjacent sites. The question is whether this energy cost can be
fully compensated by delocalization, as was the case for the HR state.
Moreover, on the torus, defects such as the above could only occur in pairs.
Assuming, then, that there is some contact energy when two such defects are in proximity,
\textit{unlike} the case for a singlet pair of defects in the HR state, this could explain
why a true zero-energy state featuring such delocalized defects is only possible in the thermodynamic limit. This would explain why no exact zero mode wave functions are known featuring these 
delocalized defects, unlike in the HR case.

Alas, the above scenario does not come to pass. We will find the asymptotic energy
of defects as shown in \Eq{defect} in the TT limit using the same perturbative approach used in the preceding section. We find that, unlike in the HR case, diagonal and off-diagonal energies
for this defect are of different orders of magnitude in $x$ in the TT limit, with the positive diagonal
part dominating. We thus find the energy of such defect, and numerical calculations will show that
it is indeed the energy of the lowest excited state in the TT limit. Our analytic result will show that this
energy does not vanish in the thermodynamic limit.

\subsection{TT perturbation theory}

Equation (\ref{second quantized}) describes a center-of-mass conserving three-particle hopping process. It is useful to explicitly spell out the first few dominant processes in the TT limit:

\begin{equation}\label{TTHam}
\begin{split}
H \sim&  \sum_n \Big\{[C+1](C_n^\dagger)^3 (C_n)^3\\
+& [9C(1-2 \kappa^2/3)^2+9] e^{-2\kappa^2/3} (C_n^\dagger)^2 C_{n\pm1}^\dagger (C_n)^2 C_{n\pm1}\\
+& [6C(1-2 \kappa^2)+6] e^{-\kappa^2} (C_n^\dagger)^3 C_{n\mp1} C_n C_{n\pm1}\\
+& [9C(1-8 \kappa^2/3)(1-2 \kappa^2/3)+9] e^{-5\kappa^2/3}\\
& C_n^\dagger (C_{n\pm1}^\dagger)^2 (C_n)^2 C_{n\pm2}\\
+& [36C(1-2 \kappa^2)^2+36] e^{-2\kappa^2} C_{n\mp1}^\dagger C_{n}^\dagger C_{n\pm1}^\dagger C_{n\mp1} C_{n} C_{n\pm1}\\
+&[9C(1-8 \kappa^2/3)^2+9] e^{-8\kappa^2/3} (C_n^\dagger)^2 C_{n\pm2}^\dagger (C_n)^2 C_{n\pm2} \Big\}
\end{split}
\end{equation}

{The four diagonal} terms out of the above dominant processes penalize states having three particles in three adjacent sites. 
It is apparent how the Hamiltonian assigns an energy to configurations (030), (210), (111), and (201) 
that is large compared to (most) off-diagonal processes. A detailed analysis similar to the one carried out
in Ref. \onlinecite{ortiz} could show that any zero mode of this Hamiltonian is necessarily dominated, in the usual sense,\cite{bernevig28,bernevig29} by occupation number eigenstates free of such configurations.
This is, of course, known to be the case for the Gaffnian wave function.\cite{bernevig28,bernevig29}
This last observation is quite generally equivalent to saying that the TT limit must be free of such configurations.
In \Eq{defect}, we see that the excited state we consider has one (201) configuration.
As in the preceding section, we write
\begin{eqnarray}
H=H_0+ \lambda H_1\,,
\end{eqnarray}
where $H_0$ contains all diagonal terms, and $H_1$ contains all off-diagonal terms,
and subtleties concerning spin fluctuations are now absent.
We see from \Eq{TTHam} that $H_0$ assigns an energy of order $ e^{-8\kappa^2/3} $
to the (201) defect. For comparison, the ground-state patterns (200) and (110) have an
$H_0$ energy of $ \mathcal{O}( e^{-18\kappa^2/3})$ and $\mathcal{O} (e^{-14\kappa^2/3}) $ per
unit cell, respectively.
We know, however, that the energy associated with the (200) and (110) unit cells will cancel order by order in $x=\exp(-\kappa^2/3)$ in perturbation theory, since we know that the ground states corresponding to these respective TT limits have zero energy. Hence, we will for now be interested in terms of order $x^8$ and lower order in $x$,
and need to worry about higher order in $x$ only if cancellation is found at order $x^8$, as it did  similarly happen in the HR case.

The zeroth order (in $\lambda\equiv1$) energy of the state \eqref{defect} can be inferred from \Eq{TTHam} as
\be
   E_0=\Big[9C\big(1-\kappa^2\frac{8}{3}\big)\big(1-\kappa^2\frac{8}{3}\big)+9\Big]e^{-8\kappa^2/3} \,.
\ee
We look for corrections at second-order in $\lambda$ that are also proportional
to $x^8=e^{-8\kappa^2/3}$. We first consider diagonal processes only.
The relevant virtual transition is
\be
\begin{split}
   \dotsc 2002002002010110110110\dotsc \\
   \longrightarrow \dotsc 2002002001200110110110\dotsc
\end{split}
\ee
From this we obtain the following energy correction:
\be
E_2=\dfrac{\Big|\Big[9C\big(1-\kappa^2\frac{8}{3}\big)\big(1-\kappa^2\frac{2}{3}\big)+9\Big]e^{-5\kappa^2/3}\Big|^2}{E_0-\Big[9C\big(1-\kappa^2\frac{2}{3}\big)\big(1-\kappa^2\frac{2}{3}\big)+9\Big]e^{-2\kappa^2/3}}\,.
\ee
At the order we are interested in, it is safe to neglect $E_0$ in the denominator.
We see that this correction is of the order of $x^8$, thus of the same order as 
$E_0$ and of opposite sign. So far, this is similar to the HR case.
Unlike in the latter, however, there is no complete cancellation between the leading
orders in $x$ in $E_0$ and $E_2$.
A positive order $x^8$ energy therefore remains. It turns out that this energy dominates
contributions from any other processes at second or higher order in perturbation theory.
We have checked explicitly up to fourth-order perturbation theory that all other such processes
contribute only higher powers in $x$.
This is true for both diagonal processes {\em and } off-diagonal processes 
that effectively translate the defect. While the latter processes will certainly
delocalize the defect in exact eigenstates, they do not affect the
energy to the leading order in $x$.
Taking into account the fact that defects of the kind considered here only occur
in pairs on the torus, we have the following relation for the gap in the TT limit:
\be\label{gap}
  E_{\sf gap}\simeq 2(E_0+E_2) = \dfrac{648 C \kappa^4}{9+C(3-2\kappa^2)^2} e^{-8\kappa^2/3}\,.
\ee

\begin{figure*}[tbh]
\includegraphics[width=12cm]{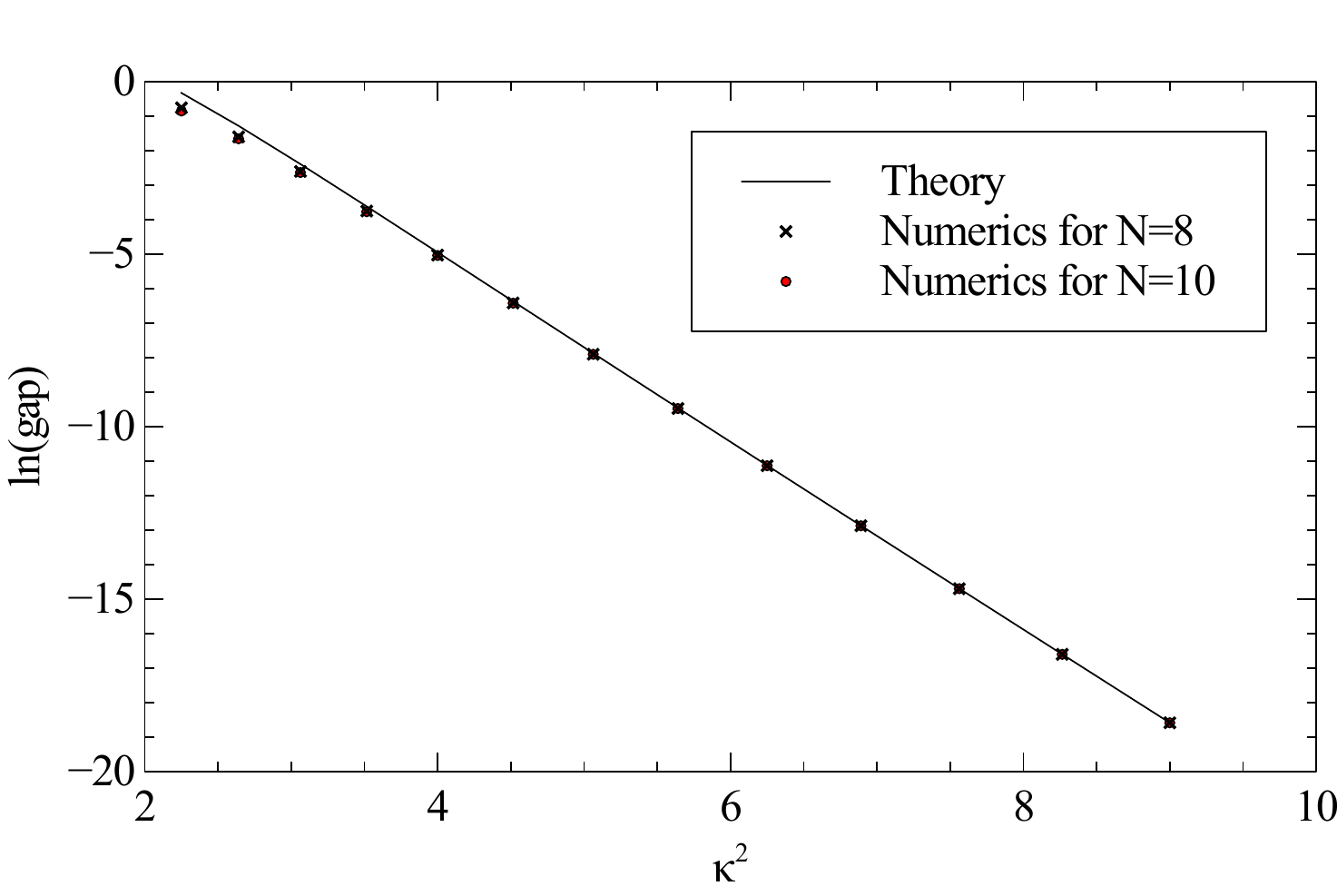}
 \caption{Comparison between the gap according to \Eq{gap} (solid line) and numerical work(dots), for $8$ and $10$ particles. The Hamiltonian parameter $C$, given by \Eq{second quantized}, has been set equal to $1$. We have obtained qualitatively similar results for different values of $C$. }
 \label{gap_fig.fig}
\end{figure*} 

\subsection{Numerics}

Equation \eqref{gap} assumes that the defect \eqref{defect} does indeed correspond to the lowest (thin torus) excitation
of the Gaffnian parent Hamiltonian \eqref{second quantized}.
In order to avoid having to consider many alternatives in the same manner, 
we compare \Eq{gap} to numerics carried out for $C=1$; see Fig. \ref{gap_fig.fig}.
The figure shows both $N=8$ and $N=10$ particle data. It is evident that there is 
small discrepancy both between the $N=8$ and the $N=10$ particle energy gap,
as well as between the latter and the prediction given by \Eq{gap}. Relative deviations
between numerical gaps and \Eq{gap} at $\kappa=3$ are $0.05\%$.
As particle number was not particularly relevant to the considerations leading to \Eq{gap},
this equation is expected to correspond to the thermodynamic (infinite cylinder) limit at {\em fixed}
but large $\kappa$ (fixed cylinder radius, small compared to a magnetic length).
The $N=8$ and $N=10$ particle data conform to this expectation.
{ Moreover,} we note that the exact first excited state  has a large overlap with the state consisting 
of the equal-amplitude superposition of all states featuring two defects of the kind shown
in  \Eq{defect}. For N=10 at $\kappa=3$, the overlap between this state and the exact first excited state is 0.999999. This confirms that we have studied the correct first excited state with our perturbative method.
We thus conclude that in the {\em $1D$} thermodynamic limit of a thin, infinite cylinder,
the Gaffnian parent Hamiltonian does not have gapless excitations.

\section{Discussion and Conclusion}\label{discussion}

The perturbative scheme used here explicitly confirms  the existence of gapless excitations in the TT limit
of the Haldane-Rezayi state. 
All of the results obtained here regarding this state had been anticipated earlier,\cite{seidel_yang}
based on the somewhat anomalous TT limit of some of the HR ground states on the torus, featuring delocalized defects. In contrast, all Gaffnian ground states have inconspicuous and simple thin torus limits. This alone could cast doubt on the existence of gapless excitations in the Gaffnian TT limit, though we have argued in Sec. \ref{gaff} that such reasoning would be naive.
Instead we have applied the same perturbative scheme employed in Sec. \ref{HR} for the HR state
to the problem of thin torus Gaffnian excitation.
We have identified certain charge-neutral defects as natural suspects for gapless excitations.
Alas, detailed calculation has shown that these excitations are gapped, and numerics strongly suggest that they are indeed the lowest excitations in the TT limit.
This implies that the 1D, thin cylinder thermodynamic limit of the Gaffnian parent Hamiltonian is gapped, unlike the similar limit for the HR parent Hamiltonian. 
This is similar to recent\cite{vaezi2014fibonacci} findings of gapped excitations in the thin torus limit 
of a ``fermionic analog'' of the Gaffnian state at filling factor $2/3$, where, however, the underlying state corresponds to a unitary CFT.
As reviewed initially, powerful arguments suggest that both Gaffnian and HR states
are gapless in the ordinary, 2D thermodynamic limit. 
On the torus, this opens up the interesting question of what happens if this 2D limit is approached asymmetrically, by first taking the 1D infinite cylinder limit at small cylinder radius, and subsequently taking the cylinder radius to infinity.
During the latter step, gapless excitations are expected to appear, under the assumption  that the 2D limit is indeed gapless. This could happen either
at a critical point at some finite cylinder radius (finite $\kappa$), or only in the limit where
the radius approaches infinity ($\kappa\rightarrow 0$). The latter is completely consistent with the idea of adiabatic continuity as a function of radius, at least for {\em any finite} radius.
For this very reason, it was cautioned in Ref. \onlinecite{seidel_yang} that  finding  
gapless excitations in the TT limit is actually a more significant indication for their existence in the
2D limit compared to the converse situation, where finding their absence in the TT limit does not necessarily imply  the existence of a gap in the 2D limit, even if adiabatic continuity is assumed.
The latter part of this cautionary remark seems to apply to the Gaffnian state.
Barring any level crossings, it is possible that the delocalized defects identified in Sec. \ref{gaff}
are adiabatically connected to gapless excitations in the 2D limit. 
This and other interesting open questions, such as the identification of the underlying cause of why gapless excitations are sometimes detectable in the TT limit and sometimes not, are left for future investigation.

\textit{Note added}. Recently we became aware of parallel work\cite{papic} by Papic, where similar conclusions are reached.

\begin{acknowledgments}
This work has been supported by the National Science
Foundation under NSF Grant No. DMR-1206781. 
\end{acknowledgments}

\bibliography{amilabib}{}
\bibliographystyle{unsrt}


\end{document}